\documentclass[preprint,proceedings]{rmaa}
\usepackage{paralist}

\usepackage{psfrag,color}

\def\gsim{\mathrel{\lower2.5pt\vbox{\lineskip=0pt\baselineskip=0pt
           \hbox{$>$}\hbox{$\sim$}}}}
\def\lsim{\mathrel{\lower2.5pt\vbox{\lineskip=0pt\baselineskip=0pt
           \hbox{$<$}\hbox{$\sim$}}}}

\def\uniden{\hbox{cm$^{-3}$}}

\def\arcmin2{\hbox{arcmin$^2$}}
\def\arcsec2{\hbox{arcsec$^2$}}

\SetYear{2002}
\SetConfTitle{Winds, Bubbles and Explosions}

\title{Line Formation in the Inner Starburst Regions of AGN} 

\author{
  Itziar Aretxaga\altaffilmark{1} 
}
\altaffiltext{1}{INAOE, Tonantzintla, Pue., Mexico.}

\shortauthor{Aretxaga I.}
\shorttitle{Line Formation in the Inner Starburst Regions of AGN}

\fulladdresses{
\item I. Aretxaga: Instituto Nacional de Astrof\'{\i}sica, \'Optica y
  Electr\'onica, Aptdo. Postal 51 y 216, 72000 Puebla, Pue., Mexico 
  (itziar@inaoep.mx).
}
\listofauthors{I. Aretxaga}
\indexauthor{Aretxaga, I.}

\abstract{
We review the evidence for young stellar populations in the inner
$\lsim 200$ pc of Active Galactic Nuclei (AGN), and the physical
mechanisms through which the stars can potentially create 
the emission lines that characterize AGN.
}

\resumen{
Revisamos la evidencia de que  existan poblaciones estelares j\'ovenes en las
regiones centrales ($\lsim 200$ pc) de los n\'ucleos gal\'acticos 
activos, y los mecanismos que los c\'umulos estelares tienen a su
alcance para crear las l\'{\i}neas de emisi\'on que caracterizan la
actividad nuclear.
}

\addkeyword{Active Galactic Nuclei}
\addkeyword{Starbursts}
\addkeyword{Supernovae}

\begin{document}
% Typeset article header
\maketitle

\section{Introduction}

The last two decades have seen a significant advance in our
understanding of the phenomenology of various classes of AGN. The
effect of the nuclear orientation on the classification of AGN has
been particularly important (see Goodrich 2001 for a review). Unified
schemes postulate that the Broad Line Region (BLR) and the continuum
emitting zone of an AGN are obscured, when observed edge-on, by a dusty
obscuring torus with a size of order $\sim$1--100~pc. Under this
scenario, and depending on the line of sight to the nucleus, identical
objects can be classified as a Seyfert 2 or Seyfert 1.

This configuration can explain several observables of Sy~2 nuclei,
for example the detection of broad lines in polarized light
(Antonucci \& Miller 1985, Miller \& Goodrich 1990) which are
directly detected in near-IR light (Goodrich et al. 1994),
the presence of kpc-scale ionizing cones
(Pogge 1988, 1989, Tadhunter \& Tsvetanov 1989), the large columns of
neutral hydrogen that absorb the X-ray emission (Mushotzky 1982, 
Bassani et al. 1999, Risaliti et al. 1999), 
and the cold far-IR colours (P\'erez Garc\'{\i}a et al. 1998). 
The scheme, taken at face value, however has some
difficulty to incorporate other observables: 
the high polarization levels observed in the
broad polarized lines of Sy~2s compared to the modest polarization of the 
continuum (Goodrich \& Miller 1989, Tran 1995, Cid Fernandes \&
Terlevich 1995), the similarity of the observed UV-continuum slopes of
Sy~1 and Sy~2s (Kinney et
al. 1991), the rich $\sim 100$~kpc environments surrounding Sy~2s
but not Sy~1s (Dultzin-Hacyan et al. 1999), the transient development 
of broad lines in otherwise classically quiescent type~2 AGN 
(Storchi-Bergmann et al. 1995, Aretxaga et al. 1999a), and the 
absence of type~2 QSOs (Hill et al. 1995).  
    
Unified schemes, nevertheless, have survived 20 years of detailed
observational tests. Sensible variations to the simple-minded
statement ``all Sy~2s are obscured Sy~1s seen edge-on'' have been
made, and the current consensus is that even if orientation is not a
unique factor in creating the variety of all AGN types, it certainly
is an important one to consider.

\section{Young stellar populations in the inner regions of AGN}

The study of the ages of the stellar populations in the inner parts of
AGN has been a source of debate over the last decade. We will define
the inner regions as those at distances $ \lsim 200$~pc from the
gravitational center of the galaxy. In most cases, these regions are
regarded as {\em nuclear} by the limitations of the resolution ($\sim
1''$) regularly achieved from ground-based optical
facilities. Powerful circumnuclear starbursts at distances of $\sim
1$~kpc from the nucleus have been known to exist in many Seyfert
galaxies for a long time (e.g. NGC~1068). The nature of the stellar
populations in the inner regions of the AGN, however, are still a
matter of active research.

In the early 90's the age determinations focused on using near-IR
absorption lines, where contamination by powerful line emission was
minimal. The absorption features due to the Ca IR triplet
$\lambda\lambda$8494,8542,8662\AA\ in nearby Sy~2 and Sy~1s were shown
to be statistically stronger than the absorptions of normal S and E
galaxies, once the possible presence of dilution by the power-law
emission of an accretion disk had been accounted for (Terlevich et
al. 1990, Jim\'enez-Benito et al. 2000, and see the figures by Nelson
\& Whittle in Terlevich 2001). This was interpreted as direct evidence
for a population of red supergiant stars that dominates the near-IR
light. The high mass-to-light ratios $L(1.6 \mu \mbox{m})/M \gsim 3
L_\odot/M_\odot$ inferred from the photospheric CO $\lambda$1.62,
$\lambda2.29\mu$m absorptions of a sample of Sy~2 nuclei
confirmed that red supergiants dominate the nuclear continuum emission
in $\sim$50\% 
of Sy~2s (Oliva et al. 1995), whilst a similar conclusion could not
be drawn for a Sy~1 sample.

With the advent of blue-sensitive detectors in many optical
facilities, by the late 90`s, the focus shifted to the detection of
the UV and Balmer absorptions of massive stars.  UV imaging and
spectroscopy of 4 Sy~2s selected by their strong [O~III]$\lambda
5007$\AA\ and 1.4GHz fluxes (which, in principle, are intrinsic AGN
properties) show a $\lsim 100-200~pc$ resolved and broken-knot
structure whose continuum spectrum is 100\%\ that of a starburst, as
derived from the strength of the absorption lines (Heckman et
al. 1997, Gonz\'alez-Delgado et al. 1998). The properties of the
starbursts are $L_{\rm SB} \approx 10^{10} - 10^{11} L_\odot$, $M_{\rm
SB} \approx 10^6 - 10^7 M_\odot$ and ages 3--6~Myr.  Even in 
the low-luminosity type~2 AGN (LINERs and Sy~2s) where the major component
of UV light is unresolved, the continuum is totally dominated by the
photospheres of OB stars (Maoz et al. 1998, Colina et al. 2002).  The
most complete surveys to date show that 50\% out of 35 Sy~2s have a
continuum which is dominated by the emission of starbursts or
post-starburst populations (Cid Fernandes et al. 2001).  A similar
study for Sy~1s has not been attempted, due to the difficulty of
decontaminating the UV--optical absorption-line spectrum from the
corresponding emission-line spectrum.

In smaller samples of radio-loud AGN a similar picture is starting to
emerge: about 40\%\ of the spectra of narrow-line radio-galaxies
show undiluted starburst features (Aretxaga et al. 2001, Tadhunter et
al. 2002), but the position of these starbursts cannot be determined
to better than $\sim$ 1 kpc from the center.

Starbursts regions confined to the inner $\sim 200$~pc can
explain some of the paradoxes that the unified scheme faces, like the
different levels of polarization of broad lines and continuum and the
similarity of the UV slopes between Sy~1 and Sy~2s, if the AGN is
not only obscured by the torus, but the torus is forming stars (Cid
Fernandes \& Terlevich 1995).

A starburst--AGN connection has been proposed in at least three
scenarios: starbursts giving birth to massive black holes
(e.g. Scoville \& Norman 1988); black holes being fed by surrounding
stellar clusters (e.g.  Perry \& Dyson 1985, see also Pittard in these
proceedings); and also pure starbursts without black holes
(e.g. Terlevich \& Melnick 1985, Terlevich et al. 1992). The evidence
for starbursts in Seyfert nuclei strongly supports some kind of
connection. However, it is still to be demonstrated that starbursts
can explain some of the phenomenology characteristic of AGN,
specifically the line emission spectrum, which ultimately is what
defines an AGN (Seyfert 1943, Baldwin et al. 1981).

\section{AGN narrow-line emission spectrum from starbursts}

Early attempts to reproduce the AGN emission-line spectrum, using stars
as the only source of ionization, have met with only moderate success
and much controversy. Terlevich \& Melnick (1985) proposed the 
existence of evolved massive stars of $T_{\rm eff} \gsim 100000$~K
(extremely-hot WC or WO stars, which they named {\em warmers}), by
directly applying the stellar evolutionary models in vogue at the
time. These stars, present in a 3--8~Myr starburst, had enough
hard-UV photons to reproduce the diagnostic lines of AGN.  It was the
addition of an optically thick stellar wind to the stellar interior
models (Maeder 1990, Schaerer et al. 1993) that suggested the
extremely-luminous blue phase did not exist, and since then,
the idea of warmers as a source of ionization in AGN has slowly
been disregarded by its proposers (Cid Fernandes 1997, Terlevich
2001).

 Normal starbursts can reproduce some of the lines observed in AGN. In
 particular, the inferred properties of the nuclear/circumnuclear
 starbursts that reproduce the observed UV absorption spectra of Sy~2s
 ($\S$2) have enough photons to ionize the whole Balmer emission
 series (Gonz\'alez-Delgado et al. 1998, Colina et al. 2002), but the
 photospheres of those young stars cannot, in general, account for the
 higher-ionization emission-line species, like [Ne~V]$\lambda
 3426$\AA\ or He~II$\lambda 4686$.

Low-ionization species, however, can be reproduced with the ionizing
power of hot stars. A starburst with normal OB stars, in dense media
($\gsim 5 \times 10^3$~cm$^{-3}$), can reproduce weak 
[O I]$\lambda 6300$\AA/H$\alpha$ LINERs
(Shields 1992, Filippenko \& Terlevich 1992). 

\section{type~IIn SNe or compact SN remnants}

\begin{figure}[!t]
  \includegraphics[width=8cm]{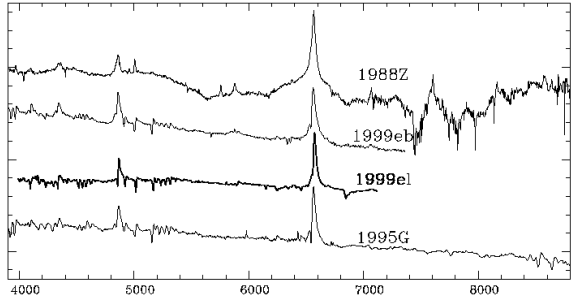}
  \caption{Rest-frame optical spectra of a collection of type IIn SNe
  (Di Carlo
  et al. 2002):  log flux (erg s$^{-1}$
  cm$^{-2}$ \AA$^{-1}$) + constant vs. wavelength (\AA).}
  \label{fig:simple}
\end{figure}

\begin{figure}[!t]
  \hspace{-0.6cm}
  \includegraphics[bb=82 69 522 676,clip,width=6.2cm, angle=90]{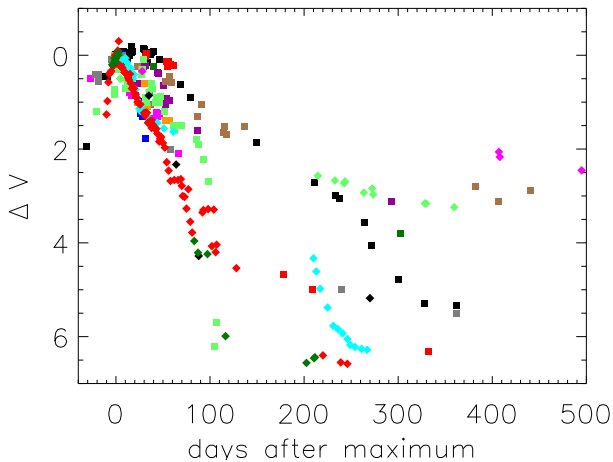}
  \caption{$V$-band light curves of 16 type~IIn SNe with known optical
    maxima, normalized to maximum light (Aretxaga et al. in prep). 
}
  \label{fig:simple}
\end{figure}

%\begin{figure}
%\plotone{}
%\caption{Rest-frame optical spectra of a collection of type IIn SNe
%  (from Di Carlo
%  et al. 2002):  log flux (erg s$^{-1}$
%  cm$^{-2}$ \AA$^{-1}$) + constant vs. wavelength (\AA).}
%\end{figure}

The modest success of using only stars to ionize the gas in an AGN,
was soon to be replaced by
the by-products of their evolution in  pure starbursts models: the
supernova explosions and the quick reprocessing of their kinetic
energy by dense circumstellar media. 
These particular SNe were first considered by Terlevich et al
in 1987 when they extrapolated the models of remnants evolving in dense
media of $\sim 10^5$~cm$^{-3}$ (Shull 1980, Wheeler et al. 1980). SNe
exploding in much denser media were soon found in the outer regions of
nearby spiral galaxies (Filippenko 1989, Stathakis \&  Sadler 1991) 
and gave rise to a different
SN spectroscopic class: type~IIn SNe (Schlegel 1990).
These objects can potentially explain
the high-ionization narrow-emission lines seen in type~2 AGN
($\S 6$), and also provide with broad-emission lines and UV--optical--IR light 
variations that resemble over long time-scales ($\gsim 1$ month) those of
type~1 AGN ($\S 5$).

The spectra of SN~IIn are characterized by the
presence of prominent narrow emission lines (hence the `n')
sitting on top of broad
components with FWHM$\lsim 15000$~km/s at maximum light
(see Fig.~1), and look extremely similar to the spectra of type 1
AGN (see a  comparison in Filippenko 1989 and Terlevich 2001). They
do not show the characteristic broad P-Cygni signatures of standard SNe,
although narrow P-Cygni profiles are detected
in some cases at high spectral resolution (e.g. SN~1997ab,
Salamanca et al. 1998).  SN~IIn
are normally associated with regions of recent star formation
(Schlegel 1990).
Despite these general characteristics, SN~IIn as a group exhibit considerable
heterogeneity (see also Filippenko 1997):
\begin{itemize}
\item some type~IIn SNe have an extremely slow decay of luminosity
after maximum light, which makes them, after 600 days, approximately 5~mag
brighter in the $V$-band than standard SN~IIP or SN~IIL (e.g. SN 1988Z,
Stathakis \& Sadler 1991, see Fig.~2); however, others have a photometric
behaviour much like standard type~IIL SNe (e.g. SN 1999el, Di Carlo et
al. 2002);
\item their peak luminosities ($M_V\sim -18.8$) 
are within the range of classical type~IIP SNe 
(Richardson et al. 2002), and thus they are not particularly
overluminous at optical wavelengths;
\item some of those SNe that decay slowly are probably hypernovae,
with kinetic energies in the range of $\sim 10^{52}$~erg (SN~1988Z,
Aretxaga et al. 1999b; SN~1997cy, Turatto et al. 2000; SN~1999E, Rigon
et al. 2002), while other slowly decaying SNe have modest integrated
energies of $\sim 10^{49}$~erg (e.g. SN~1995N, Pastorello et al., in
prep);
\item among the energetic type~IIn SNe, two are probably associated with
  gamma-ray bursts 
(SN~1997cy, Germany et al. 2000; SN~1999E, Rigon et al. 2002);
\item extremely bright radio and X-ray emission has been detected in 
some type~IIn SNe
(e.g. SN 1988Z, van Dyk et al. 1993, Fabian \& Terlevich 1995;
SN 1995N, Fox et al. 2000), but emission at these 
wavelengths is not common in others (e.g. SN~1997ab);
\item whenever the forbidden-line ratios have been used to 
estimate the density of the narrow-line producing region in type~II SNe, values
in the range $10^6 -10^9$~cm$^{-3}$ have been found 
(e.g. SN 1988Z, Stathakis \& Sadler 1991; SN 1995N, Fransson et al. 2002.;
SN 1995G, Pastorello et al. 2002).
\end{itemize}

%\begin{figure}[!t]
%  \includegraphics[width=8cm]{SED88Z.ps}
%  \caption{evol...}
%  \label{fig:simple}
%\end{figure}

It was soon recognized that the special characteristics of type IIn
SNe are due to the strong interaction of the ejecta from the explosion
with a dense circumstellar medium (Chugai 1991, Terlevich et
al. 1992), the origin of which is probably the compressed winds of the
progenitor star: slow decays and broad variable lines originate
in the dense ($\gsim10^{12}$~cm$^{-3}$) double shell structure created
by the outer and inner shocks as they sweep the dense ($\sim
10^{7}$~cm$^{-3}$) circumstellar medium and the ejecta, respectively;
and the narrow lines are produced by the unshocked circumstellar
medium which is ionized by the radiation coming from the shocks.
In the case of strong interactions, like in
SN~1988Z, where the estimated {\it radiated} energy from the radio to
X-rays in the first 10 years of evolution exceeds $3\times
10^{51}$~erg, and is probably close to  $10^{52}$~erg 
(Aretxaga et al. 1999b), i.e. two orders of magnitude larger
than typical SN events, the name `supernova' does not do justice to
the phenomenology we witness. The large radiated energies imply that
most of the kinetic energy released in the explosion must be
reprocessed into radiation within the first decade of evolution, much
as classical SN remnants behave over the course of thousands of years. 
These type~IIn SNe are referred to as
`compact supernova remnants' (cSNRs) by Terlevich
et al. (1992), which describe events where the energetics are dominated by
the conversion of kinetic into radiated energy, and not by the thermal
cooling of the expanding atmosphere of an exploding star.

\section{Type IIn SNe in the inner circumnuclear regions of AGN?}

  There is little doubt that if a SN~IIn explodes in the center of a normal
galaxy, the nucleus would be classified as a Sy~1, while
the prominent broad lines remain visible. In fact, there has been
a succession of theoretical studies that attempt to explain the phenomenology
of lines and continuum at UV to optical
wavelengths in Seyfert~1 nuclei in terms of a starburst that undergoes
SN~IIn explosions
(Terlevich et al. 1987, 1992, 1995, Aretxaga \& Terlevich 1994,
Aretxaga et al. 1997).

\begin{figure}[!t]
\vspace{-4cm}
\hspace{-1.5cm}
  \includegraphics[width=15cm]{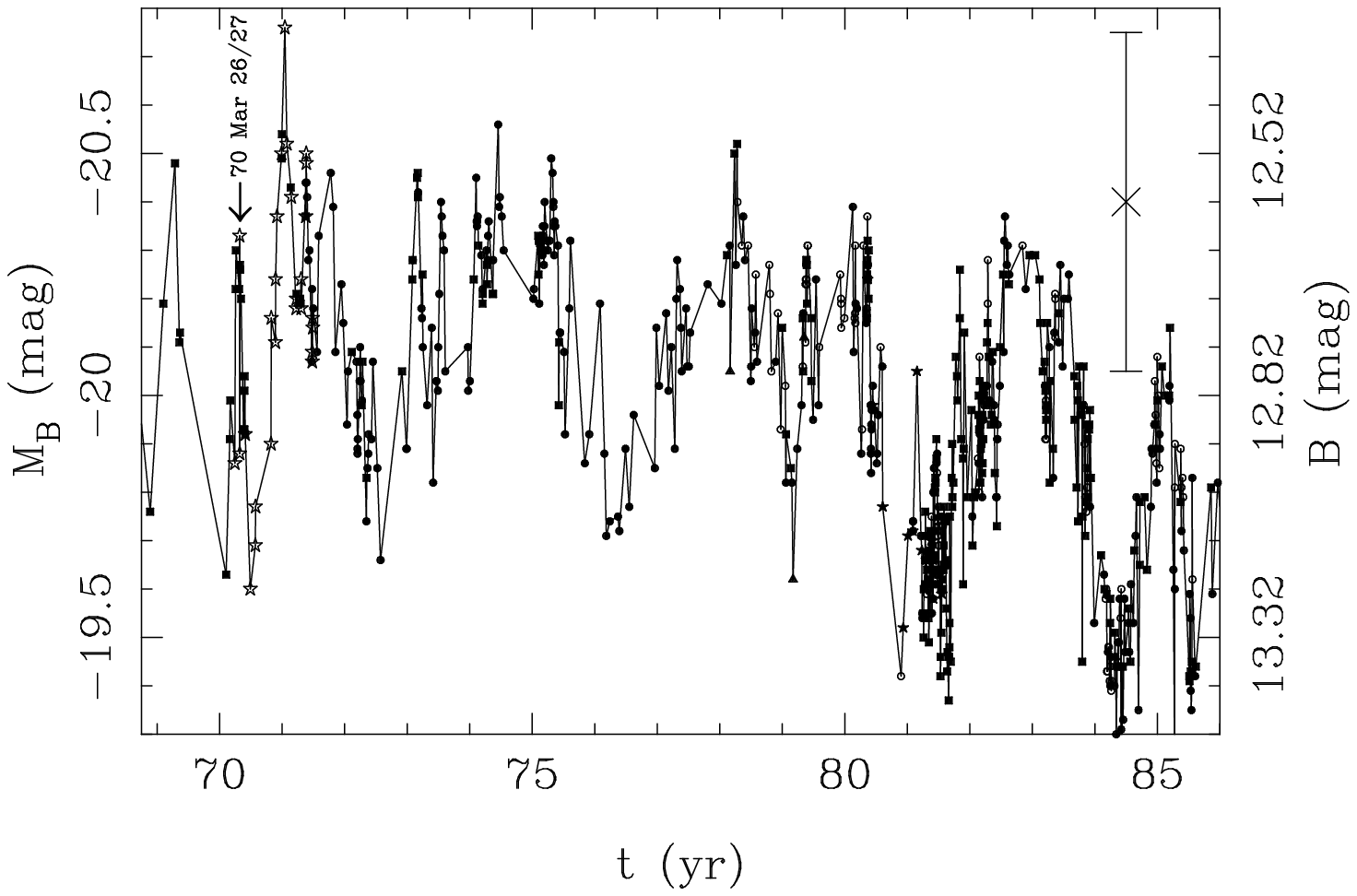}
\end{figure}
\begin{figure}[!t]
\vspace{-2cm}
\hspace{-0.3cm}
  \includegraphics[width=8.5cm]{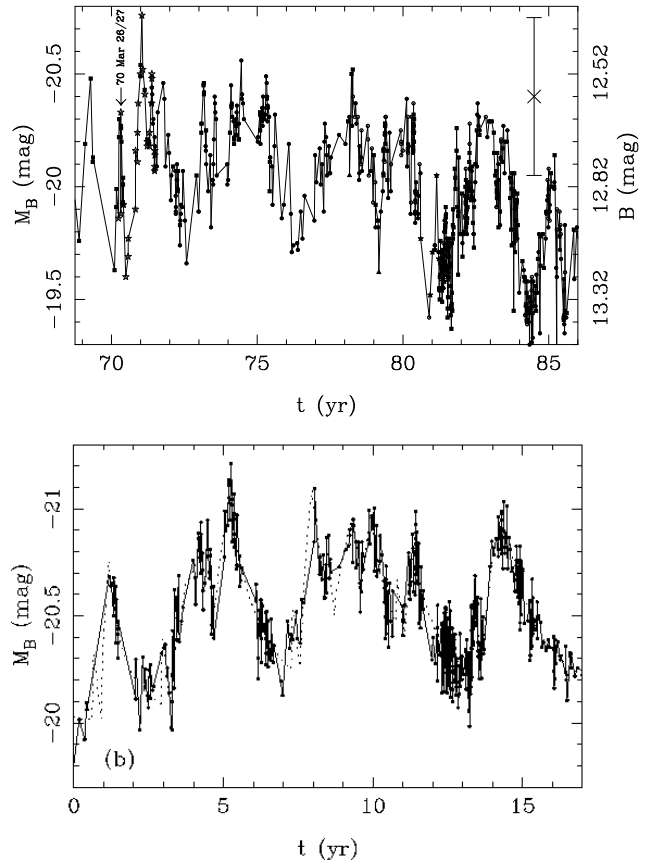}
  \caption{
{\it Top:} Light-curve of NGC~4151 in absolute $B$-band
    magnitudes. The uncertainties in the
derivation of the radial velocity of the galaxy are shown as the point with
error bars at the upper-right corner of the panel.
{\it Bottom}: Simulated light-curve for a starburst which undergoes
a SN rate $\nu_{\rm SN} = 0.3$~yr$^{-1}$,
with energy of explosions
$E = (3.0 \pm 0.1) \times 10^{51}$~erg evolving in a medium of  $n=(1.0
  \pm 0.3)\times 10^7$~\uniden. 
The sampling of points in  the simulated light-curve is identical 
to that in the top panel. 
Observational errors of $\sigma^{\rm obs}=0.1$ mag are included. The
dotted lines represent the simulated light-curve (Aretxaga \& Terlevich 1994).
}
  \label{fig:simple}
\end{figure}

Massive starbursts in the center of $\sim 50$\% of type~2 AGN have
been discovered ($\S 2$). However, it is still to be determined
whether  they also
populate the centers of type~1 AGN. The intensity of the
calcium triplet absorptions provides some evidence in this direction
(Jim\'enez-Benito et al. 2000).  

If starbursts are indeed present in the nuclei of type~1 AGN, and have
similar ages to those of type~2 AGN, they will produce a considerable
number of SNe, and {\it \underline{if}} these SNe are type~IIn's, then
they potentially can reproduce the type~1 AGN phenomenology at
IR--optical--UV wavelengths.  Since starbursts are subject to a
scaling relationship where the SN rate ($\nu_{\rm SN}$) and the
optical luminosity coming from stars ($L_B^*$) are related along the
lifetime of the SN~II explosion phase ($\sim 10-70$~Myr) by $
\nu_{\rm SN} / L_B^* \approx 2 \times 10^{-11} \rm{yr}^{-1}
L_{B\odot}^{-1} $ (Aretxaga \& Terlevich 1994), the rates required to
reproduce standard Sy~1s with pure starbursts are between 0.2 SN/yr
for AGN of luminosities close to NGC~4151 and 0.5 SN/yr for those
close to NGC~5548. These correspond to masses of the starbursts around
$10^8-10^9M_\odot$, depending on the slope and lower end of the
initial mass function. These proposed starbursts are thus 1--2 orders
of magnitude more massive than those found in Sy~2s to date. The light
curve that a cluster of luminosity similar to the nucleus of NGC~4151
would produce, if all SNe were type-IIn's, is represented in Fig.~3
together with the historically observed light curve of NGC~4151 for
comparison. The simulation, although not identical to the observed
object (since it is produced stochastically), can reproduce the basic
long-term properties such as mean luminosity, rms and power spectrum,
with just one free parameter: the density of the circumstellar medium
in which the cSNRs evolve (Aretxaga \& Terlevich 1994).

The theoretical models of cSNRs show that the UV and Balmer lines
respond to variations of the continuum on time-scales of a few days to
several tens of days, depending on the line species, and that these
lags are similar to those found in NGC~5548, but they are not created
by time  travel delays (Terlevich et
al. 1995). However, the lags have not been explored in real type~IIn
SNe since the sampling of the light variations in these sources is
still very scarce.  The only lag clearly detected is that of the
H$\alpha$ line in SN~1988Z (Turatto et al. 1993), but the lag is $\sim
200$~days, and these values are not seen (although they could
have been detected) in other type~IIn's.

The case of NGC~7582, a classical Sy 2 that suddenly mutated into a
Sy~1 (Aretxaga et al. 1999a), is probably one of the most compelling
examples where the SN explanation works, although this is by no
means a unique solution.  Many of the nuclear properties of NGC~7582
support a unified scheme where the true Sy 1 nature is hidden by an
obscuring torus: a sharp-edged [O~III] outflow in the form of a cone
is observed (Morris et al. 1985); optical spectropolarimetry does not
reveal a hidden broad-line region, but since the far-IR colours
$60\mu\mbox{m} - 25\mu$m are very red, the absence has been taken as
support for an edge-on thick torus able to block even the light
scattered towards the observer (Heisler et al. 1997); indeed a large
column density of neutral H ($N_H \sim 10^{24}$~cm$^{-2}$) also blocks
the hard X-rays (Warwick et al. 1993), but this absorption is variable and
decreased ($\Delta N_H \sim 10^{23}$~cm$^{-2}$)
at the time of the transition between Seyfert types (Turner et al. 2000). 
  The presence of stars in the nucleus is also firmly
established: Morris et al. (1985) found a steep gradient of H$\alpha$
perpendicular to the [O~III] cone, which they interpret as a 1~kpc
disk of H~II regions oriented at $60^o$ from the plane of the galaxy;
the CO absorption lines and large near-IR light-to-mass ratio are
%$L_H/M \approx 9$ solar units,
similar to those of H~II galaxies and a factor of 5 larger
than those of normal galaxies, indicating that red supergiants dominate
the light of the inner 200~pc at those wavelengths (Oliva et al. 1995); 
the expected SN rate of this starburst is 0.02 SN/yr (Aretxaga et al. 1999a).
The light variations in the optical were not consistent with a standard
reddening law variation (the Goodrich 1989 test),
which could be the result of nuclear light traveling through a 
region of less obscuration in  the torus. 
The flare, which  was followed for a few months, had a similar decaying law as
that of SN~1983K (retrospectively classified as a type~IIn, and a fast
decayer like SN~1999el) and 
line-width variations of $\sim 12000$km/s to 5000 km/s in a few 
months are also typical of
IIn's (Aretxaga et al. 1999a). The X-ray data taken during the flare
are consistent with
both explanations, either a change in reddening in the torus or a
type~IIn SN onset (Turner et al. 2000).

\section{Discussion}

Starbursts in type~2 AGN have been directly found on scales which
could correspond to the surroundings of a dusty torus.  We also could
be seeing the outskirts of a massive cluster embedded within the dusty
torus. The stars in the best studied cases (e.g. Gonz\'alez-Delgado et
al. 1998) can reproduce all the observed Balmer line emission and 
continuum, but they do not have enough ionizing photons to produce the
high-ionization lines that characterize AGN. Some type~IIn SNe provide
hard energy photons in sufficient quantities to produce the
high-ionization lines of AGN, and also coronal lines like
[Fe~X]$\lambda 6375$\AA, [Fe~VII]$\lambda 5159$\AA, [Fe~VII]$\lambda
6086$\AA\ ... (Turatto et al. 1993, Fox et al. 1995). If these SNe are
mixed with the dust, or if they are located in the central region of
the AGN and are hidden by the torus, they could reproduce the rest of the
ionization characteristics of the Sy~2s. The black hole should still be
responsible for properties such as the relativistic broad Fe~K$\alpha$
lines that are present in many Sy~2s (Turner et al. 1997), but it does
not necessarily have to be responsible for the UV--optical emission.

In turn, some type~IIn SNe in the outskirts of the  torus could be
directly visible, mimicking  the phenomenology of AGN (e.g. NGC~7582).

If one is to extrapolate this result to classical Sy~1, the masses of
the stellar clusters required are 1--2 orders of magnitude larger than
those directly detected in type~2 AGN. These could {\it per se} be
larger, or we could be seeing more of the cluster as part or all of
the dust in Sy~1s is blown up or dispersed, as suggested by
Dultzin-Hacyan et al. (1999).  Type IIn SNe do mimic many of the
characteristics of AGN, but there are still many of their properties
which are unexplored, e.g. the lags of their emission lines or the
small-scale variations.  The lag $\propto L^{1/2}$ relationship found
in AGN is one of the most outstanding predictions and successes of the
standard model of AGN (e.g. Wandel et al. 1999), where the lines
originate in (or near) an accretion disk that surrounds the
supermassive black hole.  Observationally little is known about lags
in type~IIn SNe. Only SN~1988Z shows a delay of the H$\alpha$ response
to the continuum, of $\sim 200$~days. In the rest of the SNe these
large lags are not seen, and smaller ones, if present, cannot be
characterized with the available data. A variation of the lag with the
environment of the circumstellar material is predicted, and one can
conceive of a link with the total luminosity of the cluster. The
details, however, have not been worked out, and most importantly, they
haven't been tested against bona-fide isolated SN~IIn.

It might also be the case that the stellar clusters in AGN do not
produce type~IIn SNe at all, but instead normal type~II's.  It would still be
important, however, to learn how to distinguish the optical variations
of type~IIn SNe from those of accretion processes in AGN, and estimate
the true SN rate in the clusters.

{\bf Acknowledgments:} IA's research is partly supported by CONACyT grant E--32143.

\end{document}